\documentclass{Kiso}

\usepackage[normalem]{ulem} 
\usepackage[dvips]{color} 
\renewcommand\sout{\bgroup \color{red} \ULdepth=-.5ex \ULset}

\begin{document}

\markboth{Lie-Wen Chen et al.}{Incompressibility of asymmetric
nuclear matter}

\catchline{}{}{}{}{}

\title{INCOMPRESSIBILITY OF ASYMMETRIC NUCLEAR MATTER}

\author{LIE-WEN CHEN, BAO-JUN CAI, CHUN SHEN}

\address{Department of Physics, Shanghai Jiao Tong University, Shanghai
200240, China\\
lwchen@sjtu.edu.cn}

\author{CHE MING KO and JUN XU}

\address{Cyclotron Institute and Department of Physics and Astronomy, Texas A\&M University,
College Station, TX 77843-3366, USA\\
ko@comp.tamu.edu and xujun@comp.tamu.edu}

\author{BAO-AN LI}

\address{Department of Physics and Astronomy, Texas A\&M University-Commerce, Commerce, TX
75429-3011, USA\\
Bao-An\_Li@tamu-commerce.edu}

\maketitle

\begin{history}
\received{(received date)}
\revised{(revised date)}
\end{history}

\begin{abstract}

Using an isospin- and momentum-dependent modified Gogny (MDI)
interaction, the Skyrme-Hartree-Fock (SHF) approach, and a
phenomenological modified Skyrme-like (MSL) model, we have studied
the incompressibility $K_{\mathrm{sat}}(\delta )$ of isospin
asymmetric nuclear matter at its saturation density. Our results
show that in the expansion of $K_{\mathrm{sat}}(\delta)$ in powers
of isospin asymmetry $\delta$, i.e., $K_{\mathrm{sat}}(\delta
)=K_{0}+K_{\mathrm{sat,2}}\delta^{2}+K_{\mathrm{sat,4}}\delta^{4}+O(\delta
^{6})$, the magnitude of the 4th-order $K_{\mathrm{sat,4}}$
parameter is generally small. The 2nd-order $K_{\mathrm{sat,2}}$
parameter thus essentially characterizes the isospin dependence of
the incompressibility of asymmetric nuclear matter at saturation
density. Furthermore, the $K_{\mathrm{sat,2}}$ can be expressed as
$K_{\mathrm{sat,2}}=K_{\mathrm{sym}}-6L-\frac{J_{0}}{K_{0}}L$ in
terms of the slope parameter $L$ and the curvature parameter
$K_{\mathrm{sym}}$ of the symmetry energy and the third-order
derivative parameter $J_0$ of the energy of symmetric nuclear matter
at saturation density, and we find the higher order $J_0$
contribution to $K_{\mathrm{sat,2}}$ generally cannot be neglected.
Also, we have found a linear correlation between $K_{\mathrm{sym}}$
and $L$ as well as between $J_{0}/K_{0}$ and $K_{0}$. Using these
correlations together with the empirical constraints on $K_{0}$ and
$L$, the nuclear symmetry energy $E_{\text{\textrm{sym}}}(\rho
_{0})$ at normal nuclear density, and the nucleon effective mass, we
have obtained an estimated value of $K_{\mathrm{sat,2}}=-370\pm 120$
MeV for the 2nd-order parameter in the isospin asymmetry expansion
of the incompressibility of asymmetric nuclear matter at its
saturation density.

\end{abstract}

\section{Introduction}

With the establishment or construction of many radioactive beam
facilities around the world, such as the Cooling Storage Ring (CSR)
facility at HIRFL in China \cite{CSR}, the Radioactive Ion Beam
(RIB) Factory at RIKEN in Japan \cite{Yan07}, the FAIR/GSI in
Germany \cite{FAIR}, SPIRAL2/GANIL in France \cite{SPIRAL2}, and the
Facility for Rare Isotope Beams (FRIB) in USA \cite{RIA}, it is
possible in terrestrial laboratories to explore the equation of
state (EOS) of an isospin asymmetric nuclear matter under the
extreme condition of large isospin asymmetry and thus to determine
the density dependence of the nuclear symmetry energy. This
knowledge is important for understanding not only the structure of
radioactive nuclei, the reaction dynamics induced by rare isotopes,
and the liquid-gas phase transition in asymmetric nuclear matter,
but also many critical issues in astrophysics
\cite{LiBA98,LiBA01b,Dan02a,Lat00,Lat01,Lat04,Bar05,Ste05a,CKLY07,LCK08}.

For symmetric nuclear matter with equal fractions of neutrons and
protons, its EOS is relatively well-determined. In particular, the
incompressibility of symmetric nuclear matter at its saturation
density $\rho _{0}$ has been determined to be $240\pm 20$ MeV from
analyses of the nuclear giant monopole resonances (GMR)
\cite{You99,Lui04,Ma02,Vre03,Col04,Shl06,LiT07,Gar07,Col09}, and its
EOS at densities of $2\rho _{0}<\rho <5\rho _{0}$ has also been
constrained by measurements of collective flows \cite{Dan02a} and
subthreshold kaon production \cite{Aic85,Fuc06a} in relativistic
nucleus--nucleus collisions. On the other hand, the EOS of
asymmetric nuclear matter and the density dependence of the nuclear
symmetry energy, is largely unknown except at $\rho _{0}$ where the
nuclear symmetry energy has been determined to be around $30$ MeV
from the empirical liquid-drop mass formula \cite{Mey66,Pom03}. The
values of the symmetry energy at other densities, particularly at
supra-saturation densities, are poorly known
\cite{LiBA98,LiBA01b,LCK08,Bom01,Kub03,XuC09}.

Heavy-ion collisions, especially those induced by neutron-rich
nuclei, provide a unique tool to investigate the EOS of asymmetric
nuclear matter, especially the density dependence of the nuclear
symmetry energy. During the last decade, significant progress has
been made both experimentally and theoretically on constraining the
behavior of the symmetry energy at subsaturation density using
heavy-ion reactions
\cite{Tsa04,Che05a,Che05b,LiBA05c,She07,Zha08,Tsa09,Tra09} (See Ref.
\cite{LCK08} for the most recent review). More recently, there has
also been an attempt \cite{Xiao09} to extract the symmetry energy at
supersaturation densities from the FOPI data on the $\pi ^{-}/\pi
^{+}$ ratio in central heavy-ion collisions at SIS/GSI \cite{Rei07}.
Information on the density dependence of the nuclear symmetry energy
has also been obtained from the structure of finite nuclei and their
excitations, such as the mass data \cite{Mye96}, neutron skin in
heavy nuclei \cite{Bro00}, giant dipole resonances \cite{Tri08},
pygmy dipole resonance \cite{Kli07}, and so on. These studies have
significantly improved our understanding of the EOS of isospin
asymmetric nuclear matter.

The incompressibility of asymmetric nuclear matter at its saturation
density is a basic quantity to characterize its EOS. In principle,
this information can be extracted experimentally by measuring the
GMR in neutron-rich nuclei~\cite{Bla80,Bla81}. In the present talk,
we report our recent work on the incompressibility of isospin
asymmetric nuclear matter at saturation density \cite{Che09,Che09b}.

\section{Incompressibility of isospin asymmetric nuclear matter}

\label{saturation}

The EOS of isospin asymmetric nuclear matter, given by its binding
energy per nucleon, can be generally expressed as a power series in
the isospin asymmetry $\delta =(\rho _{n}-\rho _{p})/\rho $, where
$\rho =\rho _{n}+\rho _{p}$ is the baryon density with $\rho _{n}$
and $\rho _{p}$ denoting the neutron and proton densities,
respectively. To the $4$th-order in isospin asymmetry, it is written
as
\begin{equation}
E(\rho ,\delta )=E_{0}(\rho )+E_{\mathrm{sym}}(\rho )\delta ^{2}+E_{\mathrm{%
sym,4}}(\rho )\delta ^{4}+O(\delta ^{6}),  \label{EOSANM}
\end{equation}%
where $E_{0}(\rho )=E(\rho ,\delta =0)$ is the binding energy per
nucleon of symmetric nuclear matter, and
\begin{equation}
E_{\mathrm{sym}}(\rho )=\frac{1}{2!}\frac{\partial ^{2}E(\rho ,\delta )}{%
\partial \delta ^{2}}|_{\delta =0},\text{ }E_{\mathrm{sym,4}}(\rho )=\frac{1%
}{4!}\frac{\partial ^{4}E(\rho ,\delta )}{\partial \delta
^{4}}|_{\delta =0}. \label{Esym4}
\end{equation}%
In the above, $E_{\mathrm{sym}}(\rho )$ is the\ so-called nuclear
symmetry energy and $E_{\mathrm{sym,4}}(\rho )$ is the $4$th-order
nuclear symmetry energy.

Around the normal density $\rho _{0}$, the $E_{0}(\rho )$\ can be
expanded, e.g., up to $4$th-order in density, as
\begin{equation}
E_{0}(\rho )=E_{0}(\rho _{0})+L_{0}\chi +\frac{K_{0}}{2!}\chi ^{2}+\frac{%
J_{0}}{3!}\chi ^{3}+\frac{I_{0}}{4!}\chi ^{4}+O(\chi ^{5}),
\nonumber
\end{equation}%
where $\chi $ is a dimensionless variable characterizing the
deviations of the density from the saturation density $\rho _{0}$ of
symmetric nuclear matter, and it is conventionally defined as $\chi
=\frac{\rho -\rho _{0}}{3\rho _{0}}$. The $E_{0}(\rho _{0})$ is the
binding energy per nucleon in symmetric nuclear matter at the
saturation density $\rho _{0}$ and the coefficients of other terms
are defined by
\begin{eqnarray}
L_{0} &=&3\rho _{0}\frac{dE_{0}(\rho )}{d\rho }|_{\rho =\rho
_{0}},K_{0}=9\rho _{0}^{2}\frac{d^{2}E_{0}(\rho )}{d\rho
^{2}}|_{\rho =\rho
_{0}}, \\
J_{0} &=&27\rho _{0}^{3}\frac{d^{3}E_{0}(\rho )}{d\rho ^{3}}|_{\rho
=\rho _{0}},I_{0}=81\rho _{0}^{4}\frac{d^{4}E_{0}(\rho )}{d\rho
^{4}}|_{\rho =\rho _{0}}.
\end{eqnarray}%
Obviously, we have $L_{0}=0$ according to the definition of the
saturation density $\rho _{0}$ of symmetric nuclear matter. The
coefficient $K_{0}$ is the so-called incompressibility coefficient
of symmetric nuclear matter, and it characterizes the curvature of
$E_{0}(\rho )$ at $\rho _{0}$. The coefficients $J_{0}$ and $I_{0}$
correspond, respectively, to $3$rd-order and $4$th-order
incompressibility coefficients of symmetric nuclear matter.

Around $\rho _{0}$, the $E_{\mathrm{sym}}(\rho )$\ can be similarly
expanded up to $4$th-order in $\chi $ as
\begin{equation}
E_{\mathrm{sym}}(\rho )=E_{\mathrm{sym}}(\rho _{0})+L\chi +\frac{K_{\mathrm{%
sym}}}{2!}\chi ^{2}+\frac{J_{\mathrm{sym}}}{3!}\chi ^{3}+\frac{I_{\mathrm{sym%
}}}{4!}\chi ^{4}+O(\chi ^{5}),  \label{EsymLKJI}
\end{equation}%
where $L$, $K_{\mathrm{sym}}$, $J_{\mathrm{sym}}$ and
$I_{\mathrm{sym}}$ are the slope parameter, curvature parameter,
$3$rd-order coefficient, and $4$th-order coefficient of the nuclear
symmetry energy at $\rho _{0}$. We can also expand the $4$th-order
nuclear symmetry energy $E_{\mathrm{sym,4}}(\rho )$ around
$\rho_{0}$ up to $4$th-order in $\chi $ as
\begin{equation}
E_{\mathrm{sym,4}}(\rho )=E_{\mathrm{sym,4}}(\rho _{0})+L_{\mathrm{sym,4}%
}\chi +\frac{K_{\mathrm{sym,4}}}{2}\chi ^{2}+\frac{J_{\mathrm{sym,4}}}{3!}%
\chi ^{3}+\frac{I_{\mathrm{sym,4}}}{4!}\chi ^{4}+O(\chi ^{5}),
\label{Esym4LKJI}
\end{equation}%
with $L_{\mathrm{sym,4}}$, $K_{\mathrm{sym,4}}$,
$J_{\mathrm{sym,4}}$ and $I_{\mathrm{sym,4}}$ being the slope
parameter, curvature parameter, $3$rd-order coefficient, and
$4$th-order coefficient of the $4$th-order nuclear symmetry energy
$E_{\mathrm{sym,4}}(\rho )$ at $\rho _{0}$.

Conventionally, the incompressibility of asymmetric nuclear matter
is defined at its saturation density $\rho _{\mathrm{sat}}$ where
$P(\rho ,\delta )=0$, thus also called the isobaric
incompressibility coefficient \cite{Pra85}, and is given by
\begin{equation}
K_{\mathrm{sat}}(\delta )=9\rho _{\mathrm{sat}}^{2}\frac{\partial
^{2}E(\rho ,\delta )}{\partial \rho ^{2}}|_{\rho =\rho
_{\mathrm{sat}}}. \label{KsatDef}
\end{equation}%
Its value depends on the isospin asymmetry $\delta $. Up to
$4$th-order in $\delta $, the isobaric incompressibility coefficient
$K_{\mathrm{sat}}(\delta )$ can be expressed as \cite{Che09b}
\begin{equation}
K_{\mathrm{sat}}(\delta )=K_{0}+K_{\mathrm{sat,2}}\delta ^{2}+K_{\mathrm{%
sat,4}}\delta ^{4}+O(\delta ^{6})  \label{Ksat}
\end{equation}%
with%
\begin{eqnarray}
K_{\mathrm{sat,2}} &=&K_{\mathrm{sym}}-6L-\frac{J_{0}}{K_{0}}L,
\label{Ksat2} \\
K_{\mathrm{sat,4}} &=&K_{\mathrm{sym,4}}-6L_{\mathrm{sym,4}}-\frac{J_{0}L_{%
\mathrm{sym,4}}}{K_{0}}+\frac{9L^{2}}{K_{0}}-\frac{J_{\mathrm{sym}}L}{K_{0}}
\nonumber \\
&&+\frac{I_{0}L^{2}}{2K_{0}^{2}}+\frac{J_{0}K_{\mathrm{sym}}L}{K_{0}^{2}}+%
\frac{3J_{0}L^{2}}{K_{0}^{2}}-\frac{J_{0}^{2}L^{2}}{2K_{0}^{3}}.
\label{Ksat4}
\end{eqnarray}%
It is interesting to see that with precision up to $4$th-order in
$\delta $ for the isobaric incompressibility coefficient, we only
need to know $8$ coefficients $K_{0}$, $J_{0}$, $I_{0}$, $L$,
$K_{\mathrm{sym}}$, $J_{\mathrm{sym}}$, $L_{\mathrm{sym,4}}$,
$K_{\mathrm{sym,4}}$ which are defined at the normal nuclear density
$\rho _{0}$. Furthermore, the $4$ coefficients $K_{0}$, $J_{0}$,
$L$, and $K_{\mathrm{sym}}$ determine completely the isobaric
incompressibility coefficient with precision up to $2$nd-order in
$\delta $.

It is generally believed that one can extract information on
$K_{\mathrm{sat,2}}$ by measuring the GMR in neutron-rich
nuclei~\cite{Bla80}. Usually, one can define a finite nucleus
incompressibility $K_{A}(N,Z)$ for a nucleus with $N$ neutrons and
$Z$ protons ($A=N+Z$) by the GMR energy $E_{\mathrm{GMR}}$, i.e.,
\begin{equation}
E_{\mathrm{GMR}}=\sqrt{\frac{\hbar ^{2}K_{A}(N,Z)}{m\left\langle
r^{2}\right\rangle }},
\end{equation}%
where $m$ is the nucleon mass and $\left\langle r^{2}\right\rangle $
is the mean square mass radius of the nucleus in the ground state.
Similar to the semi-empirical mass formula, the finite nucleus
incompressibility $K_{A}(N,Z) $ can be expanded as \cite{Bla80}
\begin{eqnarray}
K_{A}(N,Z)=K_{0}+K_{\mathrm{surf}}A^{-1/3}+K_{\mathrm{curv}}A^{-2/3}
&&
\nonumber \\
+(K_{\tau }+K_{\mathrm{ss}}A^{-1/3})\left( \frac{N-Z}{A}\right) ^{2} &&+K_{%
\mathrm{Coul}}\frac{Z^{2}}{A^{4/3}}+\cdot \cdot \cdot ,  \label{KA1}
\end{eqnarray}%
Neglecting the $K_{\mathrm{curv}}$ term, the $K_{\mathrm{ss}}$ term
and other higher-order terms in Eq. (\ref{KA1}), one can express the
finite nucleus incompressibility $K_{A}(N,Z)$ as
\begin{equation}
K_{A}(N,Z)=K_{0}+K_{\mathrm{surf}}A^{-1/3}+K_{\tau }\left( \frac{N-Z}{A}%
\right) ^{2}+K_{\mathrm{Coul}}\frac{Z^{2}}{A^{4/3}},  \label{KA2}
\end{equation}%
where $K_{0}$, $K_{\mathrm{surf}}$, $K_{\tau }$, and
$K_{\mathrm{coul}}$ represent the volume, surface, symmetry, and
Coulomb terms, respectively.
The $K_{\tau }$ parameter is usually thought to be equivalent to the $K_{%
\mathrm{sat,2}}$ parameter. However, we would like to stress here that the $%
K_{\mathrm{sat,2}}$ parameter is a theoretically well-defined
physical property of asymmetric nuclear matter as shown previously
while the value of the $K_{\tau }$ parameter may depend on the
details of the truncation scheme in Eq. (\ref{KA1}). As shown in
Ref.~\cite{Bla81}, $K_{\tau }$ may be related to the
isospin-dependent part of the surface properties of finite nuclei,
especially the surface symmetry energy. Therefore, cautions are
needed to interpret the $K_{\tau }$ parameter as the
$K_{\mathrm{sat,2}}$ parameter.

\section{Results and discussions}

\label{results}

In the following, we show our results based on three theoretical
models, namely, the modified finite-range Gogny effective
interaction (MDI), the Skyrme-Hartree-Fock (SHF) approach, and a
phenomenological modified Skyrme-like (MSL) model. For the details
of these models, we refer readers to Ref. \cite{Che09b}. A very
useful feature of these models is that analytical expressions for
many interesting physical quantities in asymmetric nuclear matter at
zero temperature can be obtained, and this makes the study of the
properties of asymmetric nuclear matter more transparent.

\subsection{Incompressibility of asymmetric nuclear matter at saturation density}

\begin{figure}[tbh]
\includegraphics[scale=1.05]{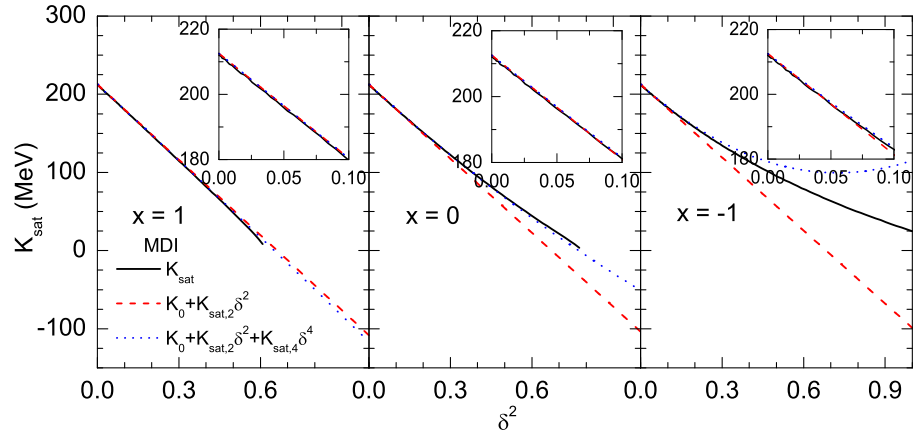}
\caption{(Color online) The incompressibility at saturation density
$K_{\mathrm{sat}}$ as a function of $\delta ^{2}$ in the MDI
interaction with $x=1$, $0$, and $-1$. Corresponding results up to
$\delta ^{2}$ and $\delta ^{4}$ are also included for comparison.
The insets display corresponding results at smaller isospin
asymmetries with $\delta ^{2}\leq 0.1$. Taken from Ref.
\protect\cite{Che09b}.} \label{KsatDel2MDI}
\end{figure}

Shown in Fig. \ref{KsatDel2MDI} is the incompressibility at
saturation density $K_{\mathrm{sat}}(\delta )$ as a function of
$\delta ^{2}$ for the MDI interaction with $x=1$, $0$, and $-1$. It
is seen that $K_{\mathrm{sat}}(\delta )$ generally decreases with
increasing isospin asymmetry. This feature is consistent with
earlier calculations based on microscopic many-body
approaches~\cite{Bom94}. Corresponding results at smaller isospin
asymmetries with $\delta ^{2}\leq 0.1$ are given in the insets of
Fig. \ref{KsatDel2MDI}, and it shows that the result from keeping up
to the $\delta ^{2}$ term approximates very well the exact
$K_{\mathrm{sat}}(\delta )$ and the contribution of
$K_{\mathrm{sat,4}}$ term is not important as
$K_{\mathrm{sat}}(\delta )$ displays a good linear correlation with
$\delta ^{2}$. This feature implies that the absolute value of
$K_{\mathrm{sat,2}}$ is much larger than that of
$K_{\mathrm{sat,4}}$. Our results are also consistent with the very
recent study based on the RMF\ model \cite{Pie09}.

\subsection{Correlations of $J_{0}$-$K_{0}$ and $L$-$K_{\mathrm{sym}}$}

As shown in Eq. (\ref{Ksat2}), the $K_{\mathrm{sat,2}}$ parameter is
completely determined by the $4$ characteristic parameters $K_{0}$,
$J_{0}$, $L$, and $K_{\mathrm{sym}}$ at the normal nuclear density.
While the $K_{0}$ parameter has been relatively well determined to
be $240\pm 20$ MeV from the nuclear GMR data
\cite{You99,Shl06,LiT07,Gar07,Col09}, the $J_{0}$ parameter is
poorly known, and there is actually no experimental information on
the $J_{0}$ parameter.

\begin{figure}[htb]
\begin{minipage}{13.5pc}
\includegraphics[scale=0.75]{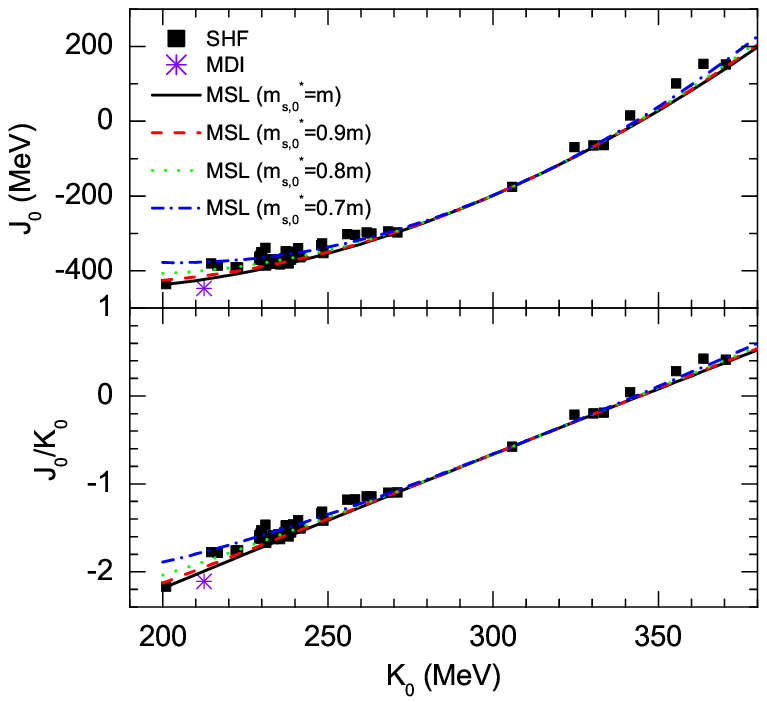}
\end{minipage}
\begin{minipage}{13.5pc}
\includegraphics[scale=0.75]{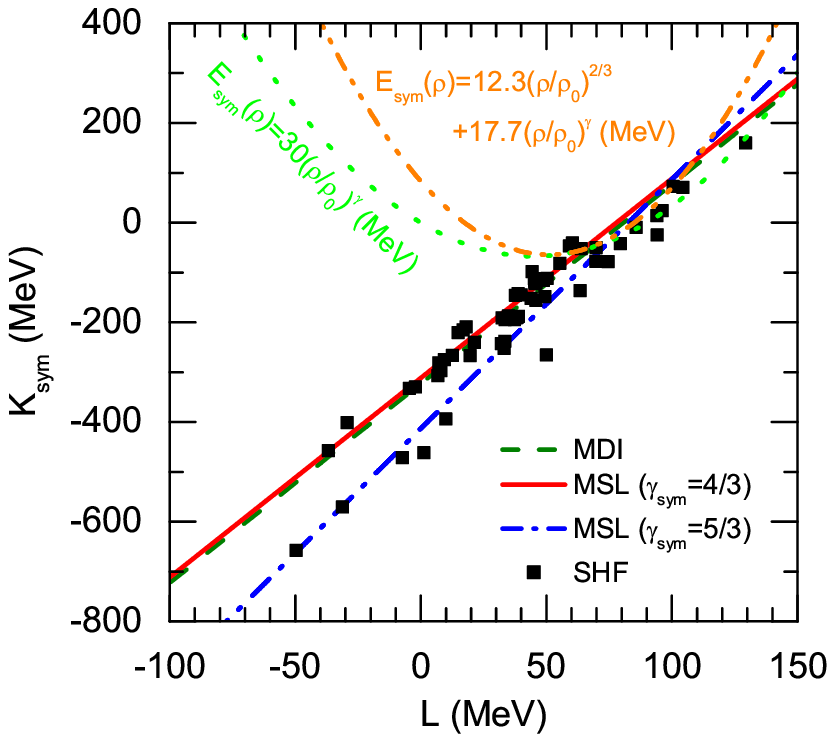}
\end{minipage}
\caption{(Color online) Left panels: The $J_{0}$ and the ratio
$J_{0}/K_{0}$ as functions of $K_{0}$ from the MSL model, the MDI
interaction and the SHF prediction with the $63$ Skyrme
interactions. Right panels: Correlation between $K_{\mathrm{sym}}$
and $L$ from the MSL model with $\gamma _{\mathrm{sym}}=4/3$ and
$5/3$, the MDI interaction and the SHF prediction with the $63$
Skyrme interactions. The results from two simple one-parameter
symmetry energies are also shown for comparison. Taken from Ref.
\protect\cite{Che09b}.} \label{Correlations}
\end{figure}

Shown in the left panels of Fig. \ref{Correlations} are $J_{0}$ and
$J_{0}/K_{0}$ as functions of $K_{0}$ in the MSL model for different
values of the nucleon effective mass $m_{s,0}^{\ast }=m$, $0.9m$,
$0.8m$ and $0.7m$. Also included in the figure are corresponding
results from the MDI interaction and the SHF prediction with $63$
Skyrme interactions (see Ref. \cite{Che09b} for the details). One
can see the $J_{0}/K_{0}$ value deviates significantly from zero for
a reasonable $K_{0}$ value, which implies that the higher order
$J_0$ contribution to $K_{\mathrm{sat,2}}$ generally cannot be
neglected. It is further seen that the correlation between $J_{0}$
and $K_{0}$ is similar among these three different models. Also, all
three models show an approximately linear correlation between
$J_{0}/K_{0}$ and $K_{0}$. We note that the correlation between
$J_{0}$ and $K_{0}$ obtained in the present work is consistent with
the early finding by Pearson \cite{Pea91}. While there do not exist
any empirical constraints on the $J_{0}$ parameter, we assume here
the correlation between $J_{0}$ and $K_{0}$ from the MSL model is
valid and then determine $J_{0}/K_{0}$ from the experimental
constraint on $K_{0}$.

Shown in the right panels of Fig. \ref{Correlations} are the
correlation between $K_{\mathrm{sym}}$ and $L$ from the MSL model
with $\gamma _{\mathrm{sym}}=4/3$ and $5/3$, the MDI interaction,
the SHF prediction with the $63$ Skyrme interactions, and two
one-parameter parameterizations on the symmetry energies. It is very
interesting to see that for larger $L$ values ($L\geq 45$ MeV which
is consistent with the constraint from heavy-ion collision data),
all above symmetry energy functionals from different models and
parameterizations give consistent predictions for the
$K_{\mathrm{sym}}$-$L$ correlation. This nice feature implies that
using these different models and parameterizations for the symmetry
energy will not influence significantly the determination of the
$K_{\mathrm{sat,2}}$ parameter. On the other hand, the
$K_{\mathrm{sym}}$-$L$ correlation from the two one-parameter
parameterizations on the symmetry energies deviate significantly
from the MDI, MSL and SHF predictions for small $L$ values.
Actually, the two forms of one-parameter parametrization for the
symmetry energy may be too simple to describe a softer symmetry
energy. As shown in Ref. \cite{Che05a}, the symmetry energy form
$E_{\mathrm{sym}}(\rho )\approx 31.6(\rho /\rho _{0})^{\gamma }$
cannot describe correctly the density dependence of the symmetry
energy from the MDI interaction with $x=1$ (the Gogny interaction).
Although there is no direct empirical information on the
$K_{\mathrm{sym}}$ parameter and some uncertainty on the
$K_{\mathrm{sym}}$-$L$ correlation still exist, we assume here that
the correlation between $K_{\mathrm{sym}}$ and $L$ from the MSL
model with $\gamma _{\mathrm{sym}}=4/3 $ and $5/3$ is valid and then
use the experimental constraint on $L$ to extract the value of
$K_{\mathrm{sym}}$.

\subsection{Constraining the $K_{\mathrm{sat,2%
}}$ parameter}

\begin{figure}[tbh]
\includegraphics[scale=1.1]{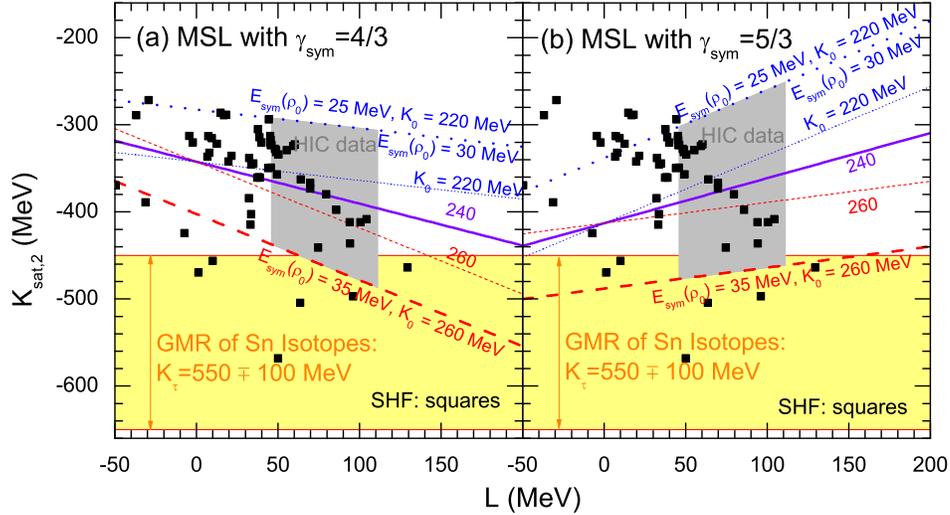}
\caption{(Color online) $K_{\mathrm{sat,2}}$ as a function of $L$
from the MSL model with $\gamma _{\mathrm{sym}}=4/3$ (a) and $5/3$
(b) and $m_{s,0}^{\ast }=0.8m$ and $m_{v,0}^{\ast }=0.7m$ for
different values of $K_{0}$ and $E_{\text{\textrm{sym}}}(\rho_{0})$.
The shaded region indicates the constraints within the MSL model
with $K_{0}=240\pm 20\ $ MeV, $E_{\text{\textrm{sym}}}(\rho
_{0})=30\pm 5$ MeV, and $46$ MeV $\leq L\leq 111$ MeV from the
heavy-ion collision data. The results from the SHF approach with 63
Skyrme interactions are also included for comparison. In addition,
the constraint of $K_{\tau }=-550\pm 100$ MeV obtained in Ref.
\protect\cite{LiT07,Gar07} from measurements of the isotopic
dependence of the GMR in even-A Sn isotopes is also indicated. Taken
from Ref. \protect\cite{Che09b}.} \label{Ksat2L}
\end{figure}
From above analyses, we can now extract information on the
$K_{\mathrm{sat,2}}$ parameter from the experimental constraints on
the $K_{0}$ parameter and the $L$ parameter. Results on
$K_{\mathrm{sat,2}}$ as a function of $L$ are shown in
Fig.\ref{Ksat2L} for $\gamma _{\mathrm{sym}}=4/3$ (panel (a)) and
$5/3$ (panel (b)) and with $K_{0}=220$, $240$, and $260$ MeV. In the
MSL model, the correlation between $K_{\mathrm{sym}}$ and $L$ also
depends on $E_{\text{\textrm{sym}}}({\rho _{0}})$ \cite{Che09b}. To
take into consideration of the uncertainty in the value of
$E_{\text{\textrm{sym}}}({\rho _{0}})$, we also include in
Fig.\ref{Ksat2L} the results with $K_{0}=220$ MeV and
$E_{\text{\textrm{sym}}}({\rho _{0}})=25 $ MeV as well as
$K_{0}=260$ MeV and $E_{\text{\textrm{sym}}}({\rho _{0}})=35 $ MeV,
which represent, respectively, the upper and lower bounds on
$K_{\mathrm{sat,2}}$ for a fixed value of $L$. The shaded region in
Fig. \ref{Ksat2L} further shows the constrained $L$ values from
heavy-ion collision data, namely, $46$ MeV $\leq L\leq 111$ MeV. The
lower limit of $L=46$ MeV is obtained from the lower bound in the
ImQMD analyses of the isospin diffusion data and the double
neutron/proton ratio \cite{Tsa09} while the upper limit of $L=111$
MeV corresponds to the upper bound of $L$ from the IBUU04 transport
model analysis of the isospin diffusion data
\cite{Tsa04,Che05a,LiBA05c,Che05b}. We note that the constraint $46$
MeV $\leq L\leq 111$ MeV obtained from heavy-ion collisions is
consistent with the analyses of the pygmy dipole resonances
\cite{Kli07}, the giant dipole resonance (GDR) of $^{208}$Pb
analyzed with Skyrme interactions \cite{Tri08}, the very precise
Thomas-Fermi model fit to the binding energies of $1654$
nuclei~\cite{Mye96}, and the recent neutron-skin analysis
\cite{Cen09}. These empirically extracted values of $L$ represent
the best and most stringent phenomenological constraints available
so far on the nuclear symmetry energy at sub-saturation densities.
It should be mentioned that the proposed experiment of
parity-violating electron scattering from $^{208}$Pb at the
Jefferson Laboratory is expected to give an independent and accurate
measurement of its neutron skin thickness (within $0.05$ fm)
\cite{Hor01,Mic05} and thus to impose a stringent constraint on the
slope parameter $L$ in future.

From the shaded region indicated in Fig. \ref{Ksat2L}, we find that
for $\gamma _{\mathrm{sym}}=4/3$, we have $-437$ MeV $\leq
K_{\mathrm{sat,2}}\leq -292$ MeV for $L=46$ MeV and $-487$ MeV $\leq
K_{\mathrm{sat,2}}\leq -306$ MeV for $L=111$ MeV. These values are
changed to $-477$ MeV $\leq K_{\mathrm{sat,2}}\leq -302$ MeV for
$L=46$ MeV and $-461$ MeV $\leq K_{\mathrm{sat,2}}\leq -251$ MeV for
$L=111$ MeV when $\gamma_{\mathrm{sym}}=5/3$ is used. Our results
thus indicate that based on the MSL model with $4/3\leq
\gamma_{\mathrm{sym}}\leq 5/3$, $K_{0}=240\pm 20$ MeV, $25$ MeV
$\leq E_{\text{\textrm{sym}}}(\rho _{0})\leq 35$ MeV,\ and $46$ MeV
$\leq L\leq 111$ MeV, the $K_{\mathrm{sat,2}}$\ parameter can vary
from $-251$ MeV to $-487$ MeV. The results shown in Fig.
\ref{Ksat2L} is obtained from a $J_{0}/K_{0}$ value that is
evaluated with the default value $m_{s,0}^{\ast }=0.8m$ in the MSL
model. Similar analyses indicate that the $K_{\mathrm{sat,2}}$\
parameter varies from $-261$ MeV to $-489$ MeV if we use
$m_{s,0}^{\ast }=0.7m$ while it varies from $-245$ MeV to $-485$ MeV
if $m_{s,0}^{\ast }=0.9m $ is used. Therefore, the extracted value
for $K_{\mathrm{sat,2}}$ is not sensitive to the variation of the
nucleon effective mass. The MSL model analyses with $4/3\leq
\gamma_{\mathrm{sym}}\leq 5/3$, $K_{0}=240\pm 20$ MeV,
$E_{\text{\textrm{sym}}}({\rho _{0}})=30\pm 5$ MeV, $46$ MeV $\leq
L\leq 111$ MeV, and $m_{s,0}^{\ast }=0.8\pm 0.1m$ thus lead to an
estimate value of $K_{\mathrm{sat,2}}=-370\pm 120$ MeV for the
2nd-order parameter in the isospin asymmetry expansion of the
incompressibility of asymmetric nuclear matter at its saturation
density.

\section{Summary and conclusions}

\label{summary}

We have studied the incompressibility of an asymmetric nuclear
matter at its saturation density. We find that in its power series
expansion in isospin asymmetry the magnitude of the 4th-order
$K_{\mathrm{sat,4}}$ parameter is generally small compared to that
of the 2nd-order $K_{\mathrm{sat,2}}$ parameter, so the latter
essentially characterizes the isospin dependence of the
incompressibility of asymmetric nuclear matter at saturation
density. Furthermore, the $K_{\mathrm{sat,2}}$ parameter can be
uniquely determined by the slope $L$ and the curvature
$K_{\mathrm{sym}}$ of nuclear symmetry energy at normal nuclear
matter density and the ratio $J_{0}/K_{0}$ of the 3rd-order and
incompressibility coefficients of symmetric nuclear matter at
saturation density, and we find the higher order $J_0$ contribution
to $K_{\mathrm{sat,2}}$ generally cannot be neglected. Since there
is no experimental information on the $J_{0}$ parameter and the
$K_{\mathrm{sym}}$ parameter, we have thus used the MSL model, which
can reasonably describe the general properties of symmetric nuclear
matter and the symmetry energy predicted by both the MDI model and
the SHF approach, to estimate the value of $K_{\mathrm{sat,2}}$.
Interestingly, we find that there exists a nicely linear correlation
between $K_{\mathrm{sym}}$ and $L$ as well as between $J_{0}/K_{0}$
and $K_{0}$ for the three different models used here, i.e., the MDI
interaction, the MSL interaction, and the SHF approach with $63$
Skyrme interactions. These correlations have allowed us to extract
the values of the $J_{0}$ parameter and the $K_{\mathrm{sym}}$
parameter from the empirical information on $K_{0}$, $L$ and
$E_{\text{\textrm{sym}}}(\rho _{0})$. In particular, using the
empirical constraints of $K_{0}=240\pm 20$ MeV,
$E_{\text{\textrm{sym}}}({\rho _{0}})=30\pm 5$ MeV, $46 $ MeV $\leq
L\leq 111$ MeV and a nucleon effective mass $m_{s,0}^{\ast }=0.8\pm
0.1m$ in the MSL model leads to an estimate of
$K_{\mathrm{sat,2}}=-370\pm 120$ MeV.

While the estimated value of $K_{\mathrm{sat,2}}=-370\pm 120$ MeV in
the present work has small overlap with the constraint of $K_{\tau
}=-550\pm 100$ MeV for the symmetry term in a semi-empirical mass
formula-like expansion of the incompressibility of finite nuclei
obtained in Refs. \cite{LiT07,Gar07} from recent measurements of the
isotopic dependence of the GMR in even-A Sn isotopes, its magnitude
is significantly smaller than that of the constrained $K_{\tau }$.
Recently, there are several studies \cite{Sag07,Pie07,Pie09} on
extracting the value of the $K_{\mathrm{sat,2}}$ parameter based on
the idea initiated by Blaizot and collaborators that the values of
both $K_{0}$ and $K_{\mathrm{sat,2}}$ should be extracted from the
same consistent theoretical model that successfully reproduces the
experimental GMR energies of a variety of nuclei. These studies show
that no single model (interaction) can simultaneously describe the
recent measurements of the isotopic dependence of the GMR in even-A
Sn isotopes and the GMR data of $^{90}$Zr and $^{208}$Pb nuclei, and
this makes it difficult to accurately determine the value of
$K_{\mathrm{sat,2}}$ from these data. Also, a very recent study
\cite{Kha09} indicates that the effect due to the nuclear
superfluidity may also affect the extraction of the
$K_{\mathrm{sat,2}}$ parameter from the nuclear GMR. As pointed out
in Ref. \cite{Pie09}, these features suggest that the $K_{\tau
}=-550\pm 100$\ MeV obtained in Ref. \cite{LiT07,Gar07} may suffer
from the same ambiguities already encountered in earlier attempts to
extract the $K_{0}$ and $K_{\mathrm{sat,2}}$ of infinite nuclear
matter from finite-nuclei extrapolations. This problem remains an
open challenge, and both experimental and theoretical insights are
needed in future studies.

\section*{Acknowledgments}
This work was supported in part by the NNSF of China under Grant
Nos. 10575071 and 10675082, MOE of China under project NCET-05-0392,
Shanghai Rising-Star Program under Grant No. 06QA14024, the SRF for
ROCS, SEM of China, the National Basic Research Program of China
(973 Program) under Contract No. 2007CB815004, the U.S. National
Science Foundation under Grant No. PHY-0652548, PHY-0757839 and
PHY-0758115, the Welch Foundation under Grant No. A-1358, the
Research Corporation under Award No. 7123 and the Texas Coordinating
Board of Higher Education Award No. 003565-0004-2007.

\end{document}